\documentstyle[referee,epsf]{mn}

\title[Transfer of Power in Non-Linear Gravitational Clustering]
      	{Transfer of Power in Non-Linear Gravitational Clustering}
\author[J.S.Bagla and T.Padmanabhan]
	{J.S.Bagla\thanks{Now at: Institute of Astronomy, Madingley Road, Cambridge CB3 0HA, U.K.  E-mail: jasjeet@ast.cam.ac.uk} and T.Padmanabhan\thanks{Email:  paddy@iucaa.ernet.in}\\
	Inter-University Centre for Astronomy and Astrophysics, Post Bag 4, 
        Ganeshkhind, Pune 411 007, INDIA}

\date{IUCAA Preprint 20/96, May 1996}

\pagerange{\pageref{firstpage}--\pageref{lastpage}}
\pubyear{1996}

\begin{document}
\label{firstpage}

\maketitle

\begin{abstract}
We investigate the transfer of power between different scales and
coupling of modes during non-linear evolution of gravitational
clustering in an expanding universe. We start with a power spectrum of
density fluctuations that is exponentially damped outside a narrow
range of scales and use numerical simulations to study evolution of
this power spectrum. Non-Linear effects generate power at other scales
with most power flowing from larger to smaller scales. The ``cascade''
of power leads to equipartition of energy at smaller scales, implying
a power spectrum with index $n\approx -1$. We find that such a
spectrum is produced in the range $1 < \delta < 200$ for density
contrast $\delta$. {\it This result continues to hold even when small
scale power is added to the initial power spectrum}. Semi-analytic
models for gravitational clustering suggest a tendency for the
effective index to move towards a critical index $n_c\approx -1$ in
this range. For $n<n_c$, power in this range grows faster than linear
rate, while if $n>n_c$, it grows at a slower rate -- thereby changing 
the index closer to $n_c$. At scales larger than the narrow range of
scales with initial power, a $k^4$ tail is produced.  We demonstrate
that non-linear small scales do not effect the growth of perturbations
at larger scales.
\end{abstract}


\section{Introduction}

Gravitational instability leads to growth of density inhomogeneities in an
expanding universe.  In Fourier space, one can study this growth as the
evolution of Fourier modes of density contrast, which evolve
independently of each other in the linear regime.  
The actual growth rate depends on the background cosmology and in a matter dominated
universe with $\Omega=1$, the amplitude of Fourier modes grows as the
expansion factor $a(t)\propto t^{2/3}$ [This is the background model
considered in this paper].  

The situation is quite different
if the linear perturbation theory is not applicable and one has to consider
the effect of coupling between different modes. The key effect of
such a coupling will be the
transfer of power between different length scales.
Some aspects of power transfer have been studied by using N-Body
simulations in the past.  
For example, Little, Weinberg and Park~\shortcite{little} conducted a
series of N-Body experiments with an initial power spectrum that was
truncated at different length scales.  They concluded that the
structure and appearance of a non-linear universe is dominated by a
small range of scales, centered around the scale that is becoming
non-linear.  A detailed study of transfer of power and evolution of
phases was carried out by Soda and Suto~\shortcite{onedph}  using
Zeldovich approximation, wherein they showed that one dimensional
collapse leads to equipartition of fluctuation power per unit
wavenumber.  They also carried out a few experiments with N-Body
simulations for gravitational clustering in three dimensions and noted
that a similar trend is indicated in this case as well.  Klypin and
Melott~\shortcite{km92} studied evolution of ratio of kinetic energy
at different scales using N-Body simulations and concluded that this
ratio tends towards a universal value in non-linear evolution of
clustering and approaches the value for the $n=-1$ power spectrum.
 
The simplest context in which one can view power transfer is the following:
Some amount of power is injected  at a given scale at $t=t_1$ and the
nonlinear coupling is allowed to transfer it to other scales.  By 
studying such an evolution numerically and computing the power spectra at 
later epochs we will be able to understand how the transfer of power
takes place. 
In this paper, we will demonstrate a few generic features of power transfer
between different modes by studying evolution of power spectrum in some
such models and  present theoretical arguments regarding their validity.
The theoretical arguments are based on some of the recent approaches to the
study of nonlinear clustering [briefly discussed in the next section] and
are presented in the spirit of a paradigm to understand the numerical
results. They should, of course, not be thought of as rigorous mathematical
proofs.

\section{Transfer of Power}

The evolution of the density contrast in Fourier space can be described by
an equation of the form \cite{pjep74}
\begin{equation}
\ddot\delta_{\bf k}+2{\dot a\over a}\dot\delta_{\bf k}=4\pi
G\bar\rho\delta_{\bf k} +Q \label{coupling}
\end{equation}
where $\delta_{\bf k}(t)$ is the Fourier transform of the density contrast,
$\bar\rho$ is the background density and $Q$ is a non-local, non-linear
function which couples the mode ${\bf k}$ to all other modes ${\bf
k'}$.  Coupling between different modes is
significant in two cases: (i) An obvious case is one with $\delta_{\bf k}
\ge 1$, i.e. the amplitude of density perturbations at the scale of interest is of
order unity or larger. (ii) A more interesting possibility arises for modes with no initial power [or exponentially small power]. In this case nonlinear
coupling provides the only driving terms, represented by $Q$ in
eqn.(\ref{coupling}). These generate power at the scale ${\bf k}$
through mode-coupling, provided initial power exists at some other scale. {\it
Note that the growth of power at the scale ${\bf k}$ will now be
governed purely by nonlinear effects even though $\delta_{\bf k} \ll
1$.} As we shall see, this fact leads to some interesting effects.  

The exact solution to eqn.(\ref{coupling}) is, of course, not known.  But
it is possible to understand some simple features of nonlinear clustering by
studying the characteristics of partial differential equations that govern
the evolution of two-point correlation function.  [See Peebles
\shortcite{lssu} and Nityananda and Padmanabhan \shortcite{rntp}.]  One may
draw the following conclusions from such a study for hierarchical models
[For more details see Padmanabhan \shortcite{tpgdeu} and
\shortcite{tpdonald}.] : 

\begin{itemize}
\item The power at a scale $x$, at an epoch $a$ is related closely to
the linearly extrapolated power, at a scale $l$ where 
\begin{equation}
l=x(1+\bar\xi(a,x))^{1/3}\,\, ; \qquad \bar\xi(a,x)\equiv{3\over x^3}
\int_0^x \xi(a,y)y^2dy \label{scales}
\end{equation}
Here $\xi$ is the two point correlation function and $a$ is the scale
factor. $\bar\xi(x,a)$ is the mean correlation averaged up to the scale
$x$ at the epoch $a$ \cite{hamilref}.

\item N-Body simulations of hierarchical models show that the
non-linear mean correlation function at $x$ relates to 
the linear mean correlation function at $l$ in an almost universal
manner \cite{hamilref}.  The relation can be approximated
by the following map \cite{potevol} 
\begin{equation}
\bar \xi (a,x) = \left\{ \hbox{ } \begin{array}{ll}
\bar \xi_L (a,l) &  \hbox{$(\bar \xi_L<1.2, \, \bar \xi<1.2)$} \\ 
0.7 \bar \xi_L(a,l)^3 & \hbox{$(1.2<\bar\xi_L<6.5, \, 1.2<\bar \xi<195)$} \\
11.7 {\bar \xi_L(a,l)}^{3/2} & \hbox{$(6.5<\bar\xi_L, \, 195<\bar \xi)$} \\
\end{array} \right.  \label{hamilton}
\end{equation}
provided we can assume stable clustering of virialised objects in the
non-linear regime.
\end{itemize}

Relation between  $\bar\xi(a,x)$ and $\bar\xi_L(a,l)$ was originally
given by Hamilton et al. \shortcite{hamilref} based on the N-body
simulation data.  The eqn.(\ref{hamilton}) is an approximate fit to
N-Body data which has the advantage of being a piecewise power law fit
in the three regimes.  The three regimes used in this power law fit
have some physical relevance as has been shown in a recent
paper \cite{tpgdeu} that also justifies the slopes of the power laws
used in each regime.   Several authors have provided more exact
fitting functions to describe the three regimes in a unified manner.  [See
Hamilton et al. \shortcite{hamilref}; Peacock and Dodds
\shortcite{peacockdodds96}; Jain, Mo and White \shortcite{jmw95}.]  There
is also some controversy regarding the actual index in the intermediate
regime [$\bar\xi \propto \bar\xi_L^3$] and whether the relation 
given above is truly ``universal" [see for example Peacock and Dodds
\shortcite{peacockdodds96}; Jain, Mo and White \shortcite{jmw95};
Padmanabhan et al. \shortcite{tpcen}.].  However, all the models suggested
in the literature lead to the following feature : In the quasi-linear
regime, there exists a critical index $n_c$ such that the growth of power
is faster than $a^2$ for spectra with $n<n_c$ and slower than $a^2$ for
spectra with $n>n_c$.

With the simple scaling given above, it is easy to show that $n_c=-1$;
but if a more accurate fitting function is used then this value may vary
around $-1$. For the sake of illustration we shall take $n_c=-1$ but
the general arguments in this paper do not depend on the particular
choice of $n_c$. The existence of such an index suggests that, during the evolution, there will be a tendency for the power spectra to acquire such an index. This conclusion - which is a direct consequence of the results
described above - is worth testing in numerical simulations. 

\section{Numerical Experiments}

We shall now try to see what these results might imply for the transfer of
power in a more general context. But before describing the results of such an experiment, we would like to spell out the theoretical expectations. this
is important because it shows - up front - what the results should be and
helps us to understand them. Needless to say, the real justification for the claims made below comes from the results of the simulations.

To begin with, it is well known that the power transfer in
gravitational clustering is mostly from large scales to small scales.
[A significant exception is the
generation of the $k^4$ tail which we shall discuss towards the end of
the paper.]
This is clearly borne out by the numerical experiments of Little,
Wienberg and Park \shortcite{little} that showed that the structure and
appearance of nonlinear structures in simulations is largely
independent of the initial power at small scales.  
Suppose we start with a power spectrum that is centered at some scale
$k_0=2\pi/L_0$ and has a small width $\Delta k$. First structures to
form in such a system are voids with a typical diameter
$L_0$. Formation and fragmentation of sheets bounding the voids leads
to generation of power at scales $L<L_0$. First bound structures form
at the mass scale corresponding to $L_0$. In such a model the linear
$\bar\xi$ at $L<L_0$  is nearly constant with an effective index of
$n\approx -3$. Assuming we can use eqn.(\ref{hamilton}) with the
local index in this case, we expect the power to grow very rapidly
as compared to the linear rate of $a^2$. [The rate of growth is $a^6$
for $n= -3$ and $a^4$ for $n=-2.5$.] Different rate of growth for
regions with different local index will lead to steepening of
the power spectrum with an accompanying slowing down of the rate of
growth. This rapid growth is expected in the quasi-linear regime and
should lead to a power spectrum with the critical index $n_c$ as its
slope.

Consider next a more complex situation, with initial power
concentrated around two scales $L_0$ and $L_1<L_0$.  If we assume that
the power at $L_1$ has a higher amplitude 
then the smaller scale(s) will reach the quasi-linear phase before the
larger one(s) and -- in the subsequent evolution -- will approach the
critical index $n_c$. We again expect the spectrum to have $n=n_c$ at
scale $L<L_1$.  If the largest non-linear scales dominate the non-linear
evolution of perturbations, as shown in a limited context
\cite{little}, then the spectrum at scales $L_1<L<L_0$ should also
approach one 
with an index $n=n_c$ {\it after the larger scale becomes 
nonlinear}.  However, if the extra small scale power implies a much
steeper spectrum, we should expect the small scale power to
grow at a {\it slower} rate in order to match up with the $n=n_c$
power spectrum at larger scales. If this is indeed the case, then the
power spectrum is effectively driven by the largest scale which is
entering the quasi-linear phase at the epoch of consideration. This
scale is expected to influence the scales in the quasi-linear regime. 

The question we would like to address in this paper is : To what
extent is the above qualitative picture supported by numerical simulations ?
To answer this question we will use N-Body simulations and study
evolution of some toy models.  We begin with a brief summary of the
N-Body code used for these numerical experiments.  For greater details
of the PM code used here see Bagla and Padmanabhan~(1996)

All simulations used a PM [Particle Mesh] code and $(128)^3$ particles in a
$(128)^3$ box.  In the units of length used here, each side of the
simulation box measures $128$ units. We used the TSC [Triangular Shaped
Cloud] for interpolation and the ``poor man's'' Poisson solver for solving
the Poisson equation in Fourier space. Force was computed in Fourier space
from the potential. For more details on PM codes, see Hockney and Eastwood
\shortcite{simbook}.
Numerical integration of the equation of motion was done using the standard
Leap-Frog method.  Step size for integration was chosen by enforcing an upper 
limit on the maximum displacement of a particle in one step.  The
``woe factor" [$10 \Delta{t}/t_{dyn}$] for simulations used here is
about $0.2$. 

In all the figures for power spectrum in this paper, we shall show only
that region in the $k$-space where the uncertainty is small. The smallest
scale shown in these figures is $1.5 L_{ny}$ where $L_{ny}$ is the Nyquist
scale.  Average number of particles enclosed in a cube of this size, or
larger, contains sufficiently large number of particles and hence the error
in power spectrum is fairly small at these scales.  The largest scale used
in these plots is $L=L_{box}/2$, thus the averaging for power spectrum is
done over at least eight independent regions in the simulation box.

We used three models for our study.  Parameters of these models were chosen
as follows

\begin{itemize}
\item Initial power spectrum for model I, the ``reference" model, was
a Gaussian peaked at the scale $k_0=2\pi/L_0 ; L_0=24$ and having a
spread $\Delta k=2\pi/128$. The amplitude of the peak was chosen so
that $\Delta_{lin} (k_0=2\pi /L_0, a=0.25)=1$, where $\Delta^2(k)=k^3
P(k)/(2\pi^2)$ and $P(k)$ is the power spectrum. Needless to say, the
simulation starts while the peak of the Gaussian is in the linear
regime $(\Delta(k_0) \ll 1)$. 

\item Model II had initial power concentrated in two narrow windows in
$k$-space. In addition to power around $L_0=24$ as in model I, we
added power at $k_1=2\pi/L_1 ; L_1=8$ using a Gaussian with same width
as that used in model I. Amplitude at $L_1$ was chosen five times
higher than that at $L_0=24$, thus $\Delta_{lin} (k_1,a=0.05)=1$.

\item Model III was similar to model II, with the small scale peak
shifted to $k_1=2\pi/L_1 ; L_1=12$. The amplitude of the small scale
peak was the same as in Model II. 
\end{itemize}

\begin{figure}
\epsfxsize=6.3 true in \epsfbox[52 73 560 300]{fig1.1.1.ps}
\epsfxsize=6.3 true in \epsfbox[52 73 560 300]{fig1.1.2.ps}
\caption{This figure shows a slice from simulations of model~I at four
epochs.  The upper left panel corresponds to $a=0.25$, the upper right
panel to $a=0.5$, the lower left panel to $a=1$ and the lower right
panel to $a=2$.  The thickness of the slice is $6L$ and the width and
height are $128L$ where $L$ is one grid length.} 
\end{figure}
\setcounter{figure}{0}
\begin{figure}
\epsfxsize=6.3 true in \epsfbox[52 73 560 300]{fig1.2.1.ps}
\epsfxsize=6.3 true in \epsfbox[52 73 560 300]{fig1.2.2.ps}
\caption{Continued.  This figure shows the corresponding slice from
simulations of model~II.}
\end{figure}
\setcounter{figure}{0}
\begin{figure}
\epsfxsize=6.3 true in \epsfbox[52 73 560 300]{fig1.3.1.ps}
\epsfxsize=6.3 true in \epsfbox[52 73 560 300]{fig1.3.2.ps}
\caption{Continued.  This figure shows the corresponding slice from
simulations of model~III.}
\end{figure}

\begin{figure}
\epsfxsize=5.7 truein\epsfbox[28 17 550 359]{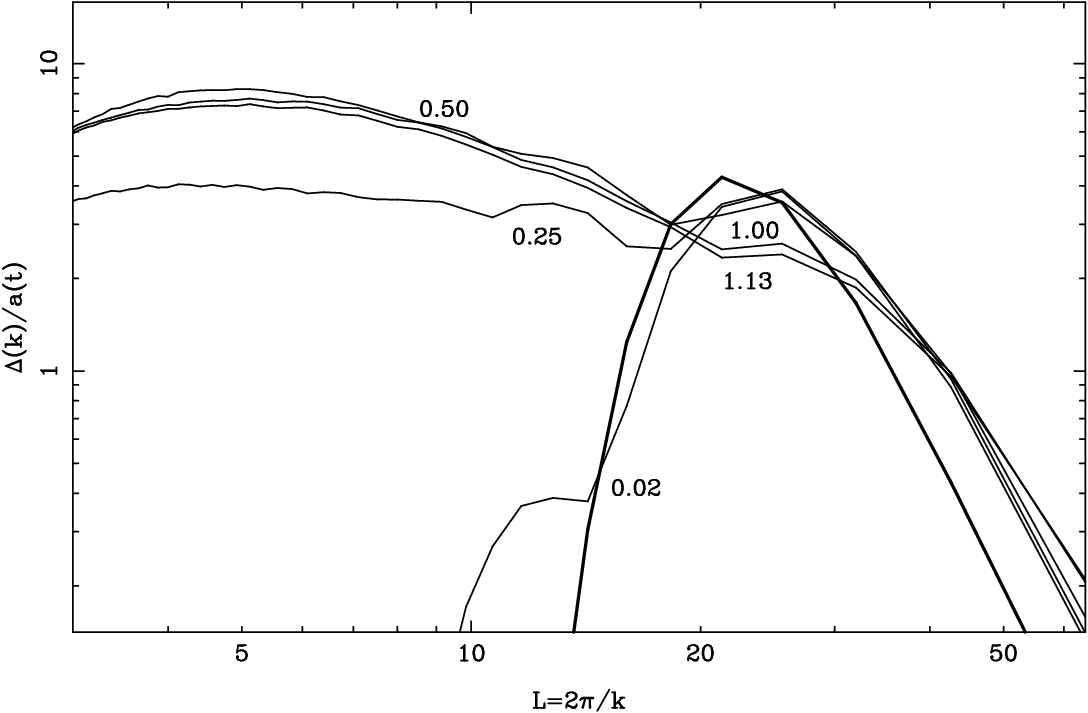}
\epsfxsize=5.7 truein\epsfbox[28 17 550 359]{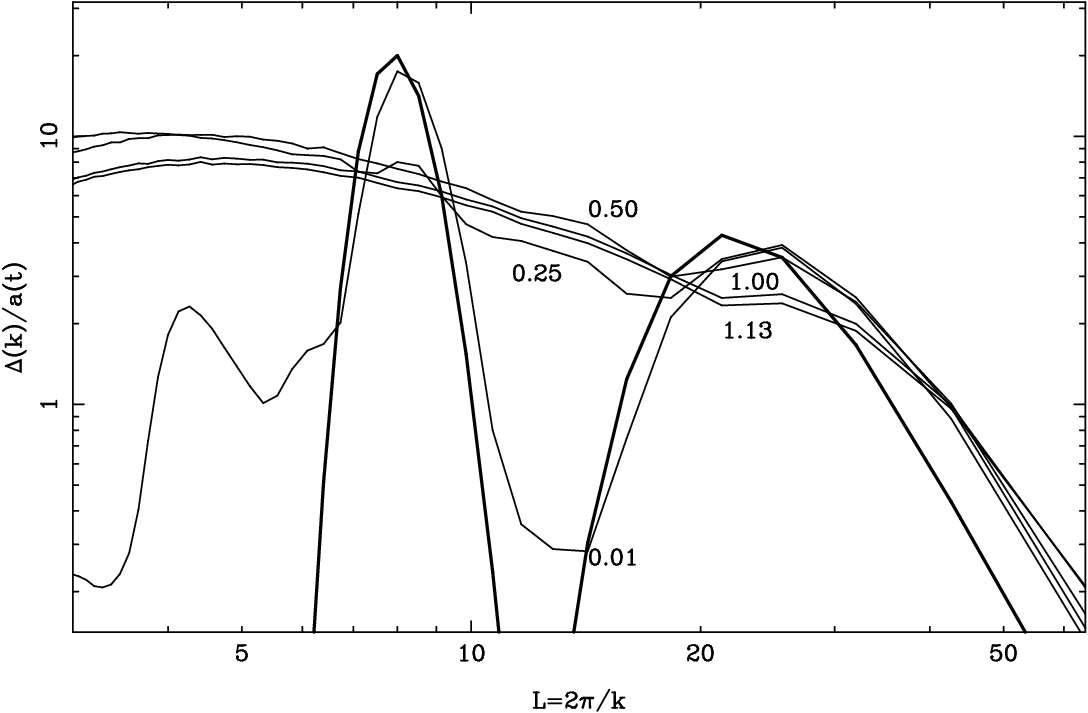}
\caption{This figure shows evolution of power spectra for the three
models. Thick lines show the linear power spectrum for these models
and thin lines show the non-linear power spectrum from N-Body
simulations. The y-axis is square root of power per logarithmic scale
divided by $a$. With linear growth rate divided out, only non-linear
evolution can modify the spectrum in this plot. X-axis is the length
scale. In model I (top panel) the initial power
spectrum is a Gaussian peaked at a length scale of $L_0=24$ and
amplitude adjusted to make it reach non-linearity at $a=0.25$. Power
spectra for different epochs are labeled by the scale factor. This
figure demonstrates that power is generated at smaller scales and at
late times this power saturates to a power spectrum with $n\approx
-1$. The second panel shows the evolution of model II. In this 
model there is additional power (apart from what is there in model I)
at $L_1=8$ with an amplitude five times higher.} 
\end{figure}
\setcounter{figure}{1}
\begin{figure}
\epsfxsize=5.7 truein\epsfbox[28 17 550 359]{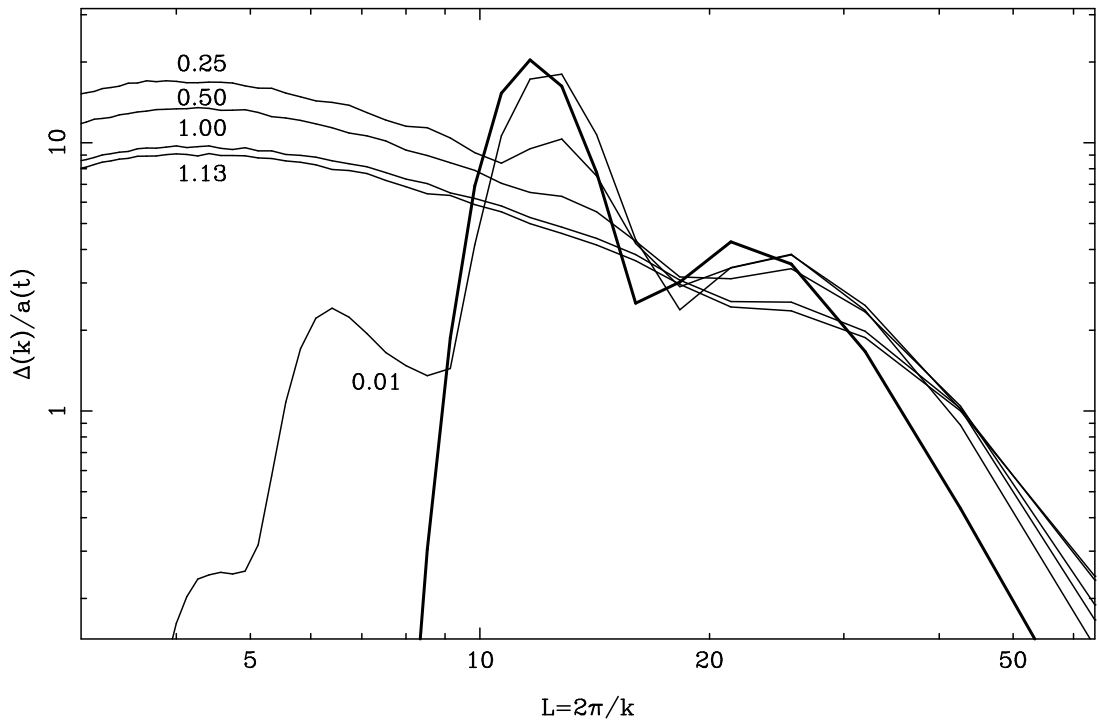}
\caption{Continued.  This panel shows
the same curves for model III. In this 
model there is additional power (apart from what is there in model I)
at $L_1=12$ with an amplitude five times higher. Notice that at late
times power spectra of all three models are very similar.}
\end{figure}

We now describe results of simulations of these models.  Panels of
figure~1 show a slice from the simulation volume at four epochs for
each model.  A comparison of non-linear structures in these panels
shows that at early epochs the appearance of non-linear objects in these
models is very different from each other.  Dissimilarities between
different models slowly disappear and the appearance at later epochs is
dominated by the larger wave-mode.  This is true even for model~III that
has a lot of small 
scale power.  [If we define an effective index of the power spectrum
by joining the peaks of two Gaussian-s then the effective index
$n_{eff} \sim 0$ for model~II and $n_{eff} \sim 2$ for model~III.]
It is clear from these pictures that the large scale mode takes a
longer time to assert itself for the case with greater small scale
power.  

To compare the evolution of these models in a more quantitative
fashion we use the power spectrum.  Top panel of figure~2
shows evolution of power spectrum for model~I. The y-axis is
$\Delta(k)/a$, the power per logarithmic scale divided by the linear
growth factor. This is plotted as a function of scale $L=2\pi/k$ for
different values of scale factor $a(t)$, curves are labeled by the
value of $a$. As we have divided the power spectrum by its linear rate
of growth, the change of shape of the spectrum occurs strictly because
of non-linear mode coupling. It is clear from this figure that power at
small scales grows rapidly and saturates to growth at a rate close to
the linear rate [shown by crowding of curves] at later epochs. The
effective index for the power spectrum approaches $n=-1$ within the
accuracy of the simulations. Thus this figure clearly demonstrates the
general features we expected from our understanding of scaling
relations.

The other two panels of figure~2 show the corresponding curves for
models~II and III respectively. These models had power at small scales
in addition to the power at large scales. The large scale power is
same in all the models and the initial conditions for the relevant
region in $k$-space had same initial phases, making comparison
meaningful. These figures show that when large scales become
non-linear, power at small scales grows at a rate {\it slower} than the
linear rate  in the quasi-linear regime, till we get a power spectrum
with $n\approx -1$ for $L<L_0$. {\it The amplitude of power spectrum
for models II and III at this stage is same as the corresponding power
spectrum in model I}. A comparison of the lower panels also shows that
the approach to $n_c$ is slower for the model with more small
scale power (model III).

\begin{figure}
\epsfxsize=5.7 truein\epsfbox[34 22 550 359]{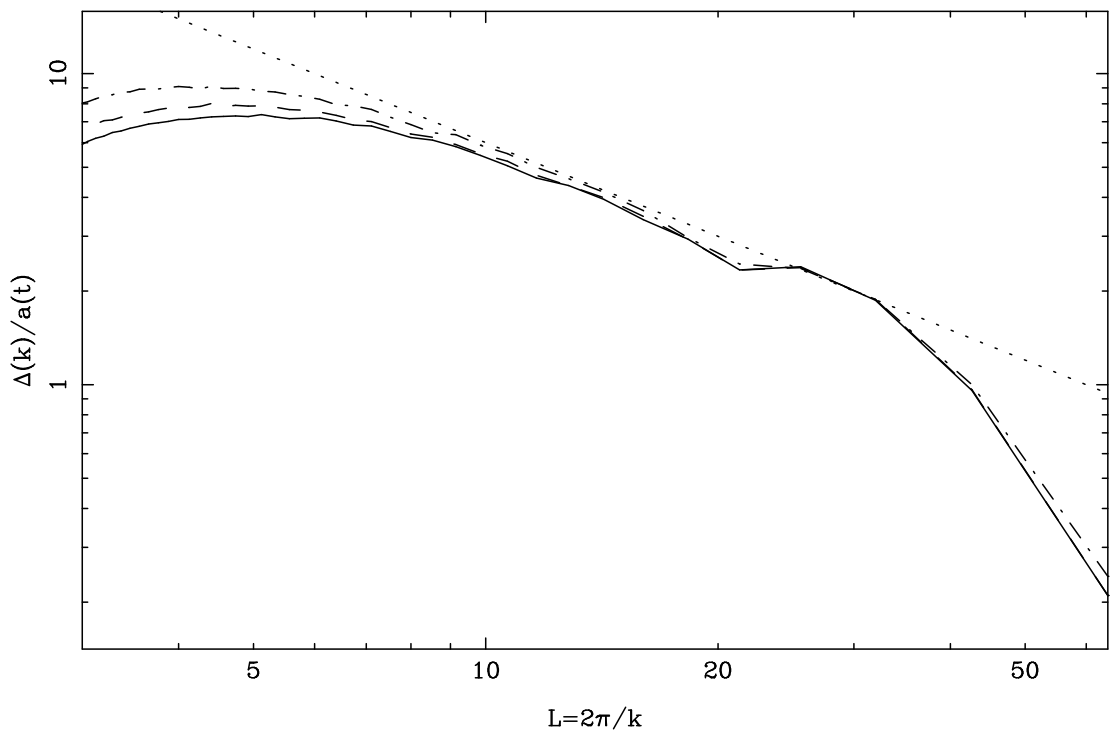}
\caption{Comparison of power spectra for the three models at a late
epoch. This figure shows that non-linear evolution has erased large
differences between model I (thick line), model II (dashed line) and
model III (dot-dashed line). The dotted straight line corresponds to
a power spectrum with index $n=-1$.} 
\end{figure}

Figure~3 shows power spectra of all three models at a late epoch. At
this epoch $\Delta_{lin}(k_0)=4.5$ and it is clear from this figure
that the power spectra of these models are very similar to one
another. 

The growth of power at three different scales, $L=8,12,24$, is plotted
for the three models in figure~4. The thick, dashed and dot-dashed lines
represent models I, II and III respectively. Curves have been
labeled by the length scale. The thick lines demonstrate power
generated by mode coupling at the two smaller length scales. From the
dashed lines for model II, one can see that power at $L=8$ decreases
with respect to the linear rate of growth so as to asymptotically
match with the amplitude in the reference model, i.e., model I. The same
effect is seen in the dot-dashed lines for model III. This figure also
shows that the existence of power at $L=8$ does not influence
evolution of power at $L=12$, if there exists power at a larger
scale. In this sense, gravitational clustering transfers power from
larger to smaller scales [``cascades"] but not in the opposite direction
[``does not inverse cascade"]. Figure 3 and 4 {\it demonstrate} this
fact very clearly.

\begin{figure}
\epsfxsize=5.7 truein\epsfbox[34 24 543 365]{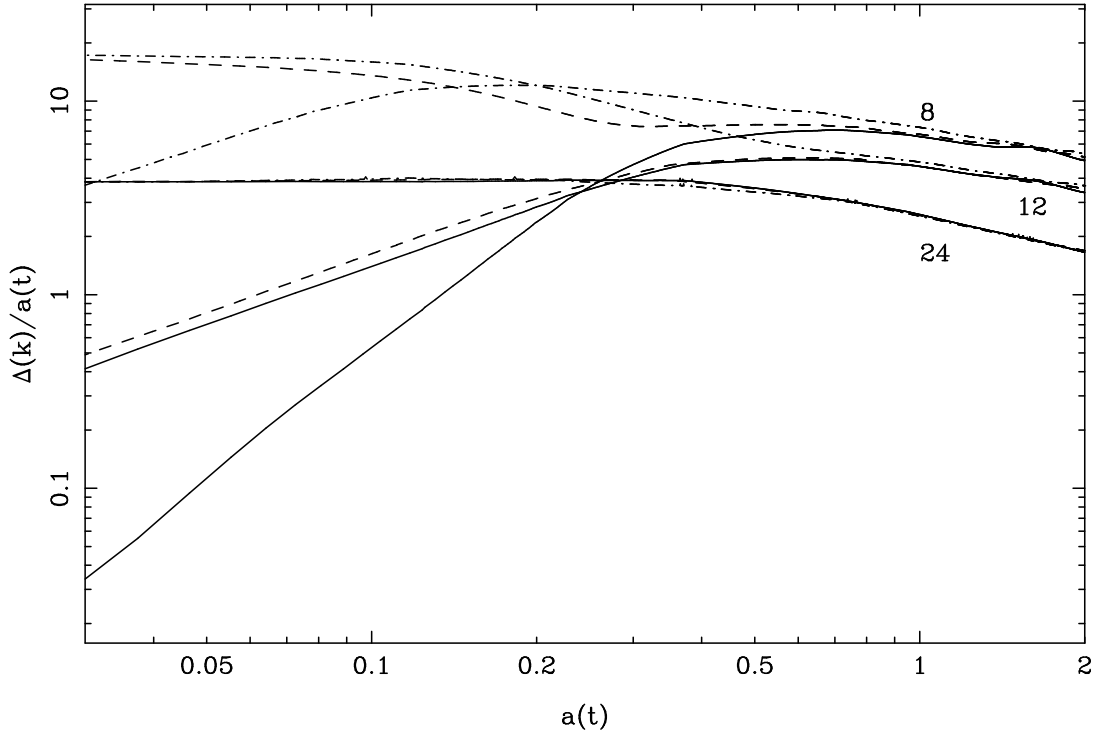}
\caption{Growth of power at three scales $L=8,12,24$
for models I (thick line), model II (dashed line) and model III
(dot-dashed line). This figure shows that at late times the power in
all three models approaches the same value. In particular, the growth
of power at $L=24$ is not influenced by additional power present at
smaller scales in model II and III, showing that non-linear small
scales do not influence larger scales.}
\end{figure}

\subsection{Critical Index}

The two panels of figure~5 illustrate two features related to the
existence of fixed points in a clear manner. In the top panel we have
plotted index of growth $n_a\equiv(\partial \ln\bar\xi(a,x)/\partial
\ln a)_x$ as a function of $\bar\xi$ in the quasi-linear regime. Curves
correspond to an input spectrum with index $n=-2,-1,1$. The dashed
horizontal line at $n_a=2$ represents the linear growth rate. An index
above the horizontal line will represent a rate of growth faster than
linear growth rate and the one below will represent a rate which is
slower than the linear rate. It is clear that -- in the quasi-linear
regime -- the curve for $n=-1$ closely follows the linear growth while
$n=-2$ grows faster and $n=1$ grows slower; so the critical index is
$n_c\approx -1$. The curves are based on the fitting formula due to
Hamilton et al \shortcite{hamilref}. Other fitting formulas suggested
by Jain, Mo and White \shortcite{jmw95} and Peacock and Dodds
\shortcite{peacockdodds96} give somewhat different curves but all
these models have fixed points close to $n_c=-1$.

The lower panel of figure~5 shows the slope $n_x = -3 - (\partial\ln
{\bar\xi} /\partial \ln{x})_a $ of $\bar\xi$ for different power law
spectra. It is clear that $n_x$ crowds around $n_c\approx -1$ in the
quasi-linear regime. 

As an aside, we can derive an upper limit on the index of an arbitrary
power spectrum.  Consider the pair conservation equation [eqn.(20) of
Nityananda and Padmanabhan~(1994)] for a set of particles.  We can
rewrite this equation by using the definition of the index of the
power spectrum $n_x$ and the index of the rate of growth $n_a$
\begin{equation}
n_a = h \left( \frac{3}{\bar\xi} - n_x \right)
\end{equation}
for the growing mode, the  amplitude of perturbations is a monotonically increasing quantity, ensuring that
 the index $n_a$ is always positive.  The pair velocity $h$
is also a positive 
quantity at scales with $\bar\xi > 0$.  Therefore the index $n_x$
satisfies the following inequality.
\begin{equation}
n_x <  \frac{3}{\bar\xi}
\end{equation}
We have plotted the curve $3/\bar\xi$ in the lower panel of figure~5
as a dotted line.  This inequality must be satisfied by non-linear
structures that have grown out of small inhomogeneities via
gravitational instability.  Any distribution of mass that does not
satisfy this inequality could not have formed due to gravitational
collapse only.  

\begin{figure}
\epsfxsize=5.7 truein\epsfbox[37 402 558 744]{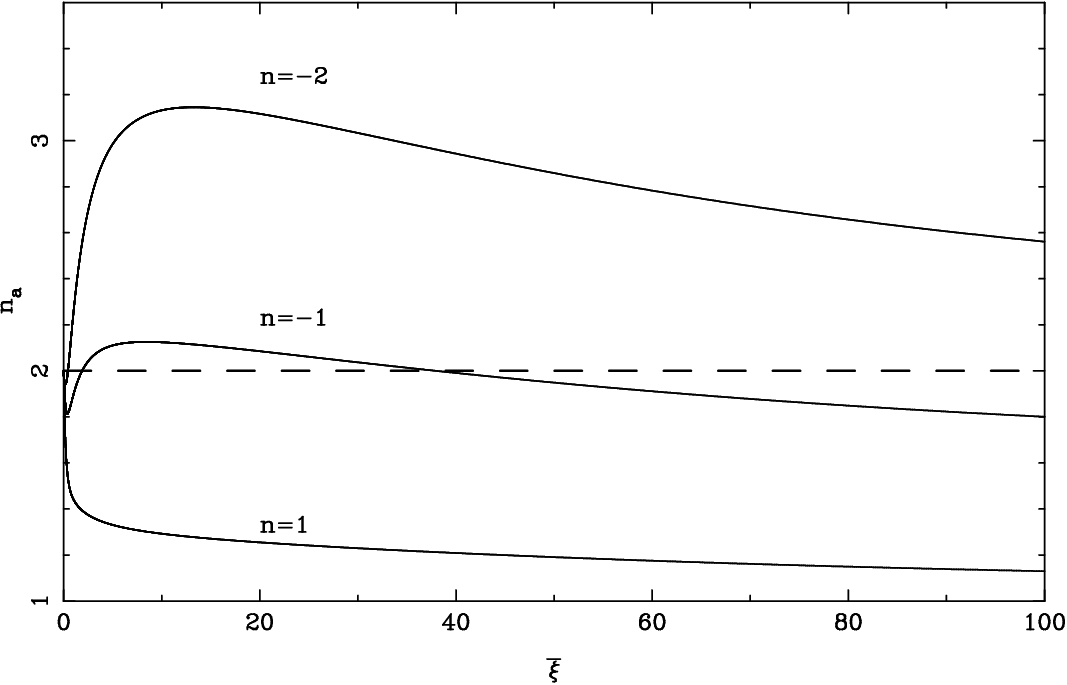}
\epsfxsize=5.7 truein\epsfbox[37 399 552 737]{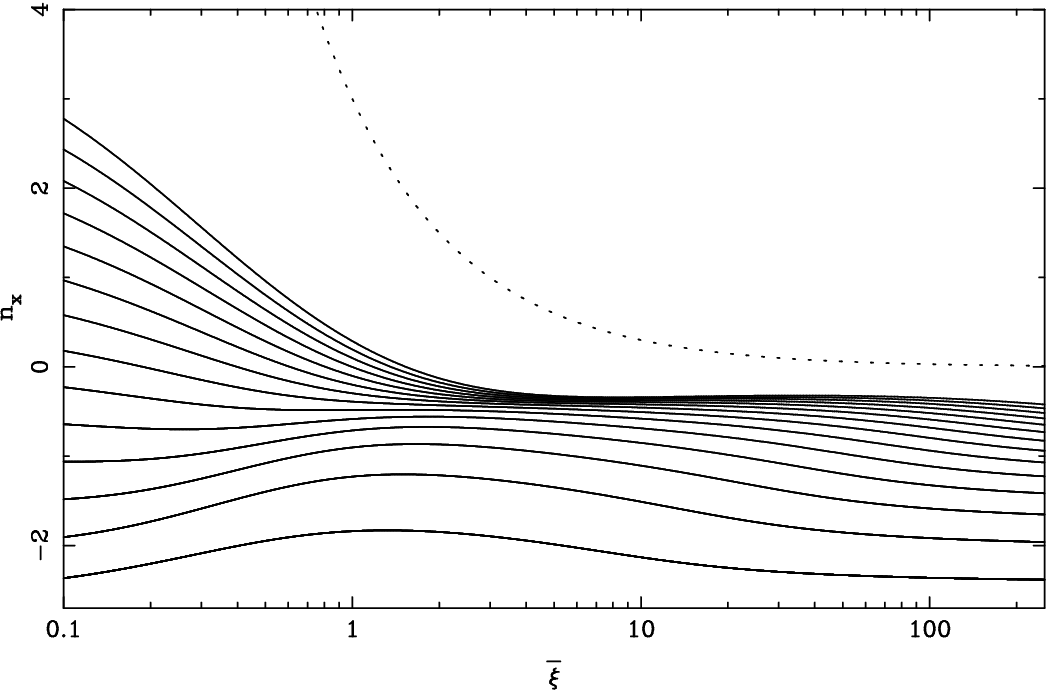}
\caption{The top panel shows exponent of rate of growth of density
fluctuations 
as a function of amplitude. We have plotted the rate of growth for
three scale invariant spectra $n=-2, -1, 1$. The dashed horizontal
line indicates the exponent for linear growth. For the range
$1<\delta<100$, the $n=-1$ spectrum grows as in linear theory; $n<-1$
grows faster and $n>-1$ grows slower. 
The lower panel shows the evolution of index
$n_x=-3-(\partial\ln {\bar\xi} /\partial \ln{x})_a$ with
$\bar\xi$. Indices vary from $n=-2.5$ to $n=4.0$ in steps of
$0.5$. The tendency for $n_x$ to crowd around $n_c=-1$ is apparent in
the quasi-linear regime. The dashed curve is a bounding curve for the
index ($n_x < 3 /\bar\xi$) if perturbations grow via gravitational
instability.}
\end{figure}

The index $n_c=-1$ corresponds to the isothermal profile with
$\bar\xi(a,x)=a^2x^{-2}$ and has some interesting features to
recommend it as a candidate for fixed point. 
For example, in the $n=-1$ spectrum each logarithmic scale contributes the same
amount of correlation potential energy. [for more details, see
Padmanabhan \shortcite{tpgdeu}.]  If the regime is modeled by scale
invariant radial flows, then the kinetic energy will scale in the same
way. It is conceivable that flow of power leads to such an
equipartition state as a fixed point though it is difficult prove such
a result in any generality.  Equipartition of kinetic energy and the
role of $n=-1$ as the transition index has been pointed out previously
by Klypin and Melott \shortcite{km92}. They studied evolution of
kinetic energy at different scales for various models to arrive at this
conclusion. It would be interesting to see whether the existence of
such a fixed point can be derived, starting from the first principles,
say, from the study of the equation (\ref{coupling}).

The eqns. \ref{hamilton} also show that, in the non-linear regime with
$\bar\xi >200$, the fixed point is $n_{c,NL}=-2$. Speculating along
similar lines, we would expect the gravitational clustering to lead to
an $x^{-1}$ profile at the non-linear end changing over to $x^{-2}$ in
the quasi-linear regime.

\begin{figure}
\epsfxsize=5.7 truein\epsfbox[34 22 550 364]{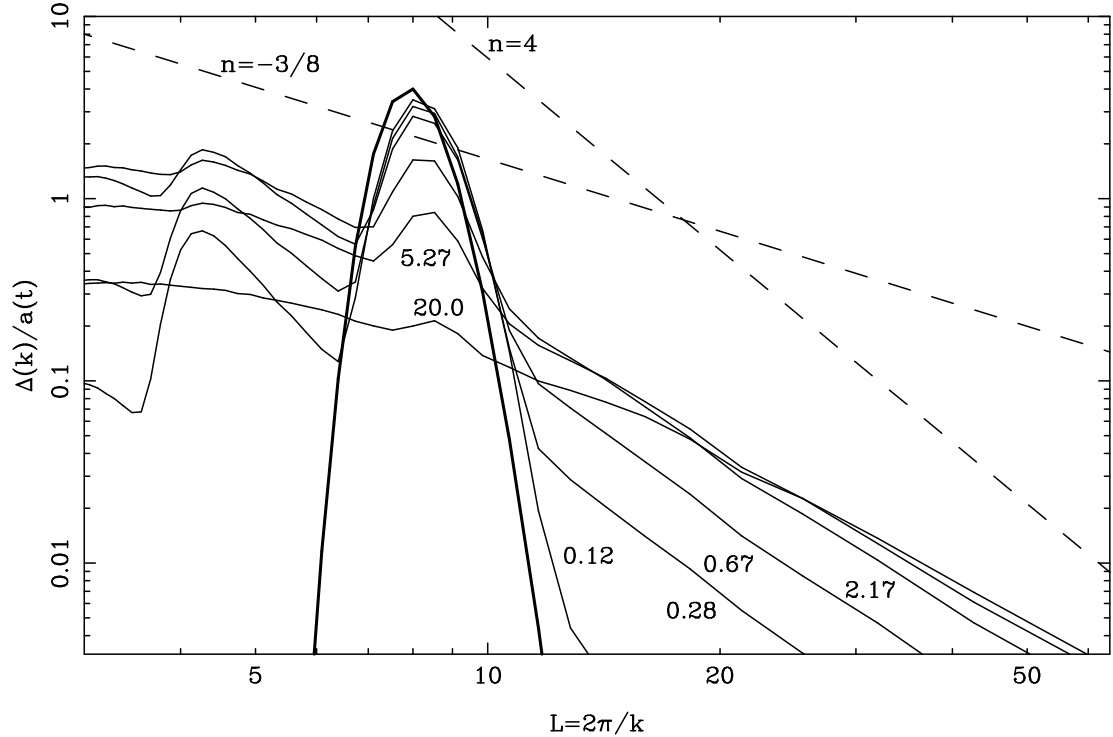}
\caption{Influence of small scales on large scales. The axes are same
as figure~2. Thick lines shows the initial power spectrum and other
curves show the non-linear spectrum at various epochs, labeled by
$a$. This figure shows generation and evolution of the $k^4$ tail. It
is seen that the slope of this tail changes rapidly and later goes
over to the expected quasi-linear index of $n=-3/8$. Reference lines
with these slopes are represented as dashed lines.}
\end{figure}

\begin{figure}
\epsfxsize = 5.7 true in \epsfbox[36 26 541 365]{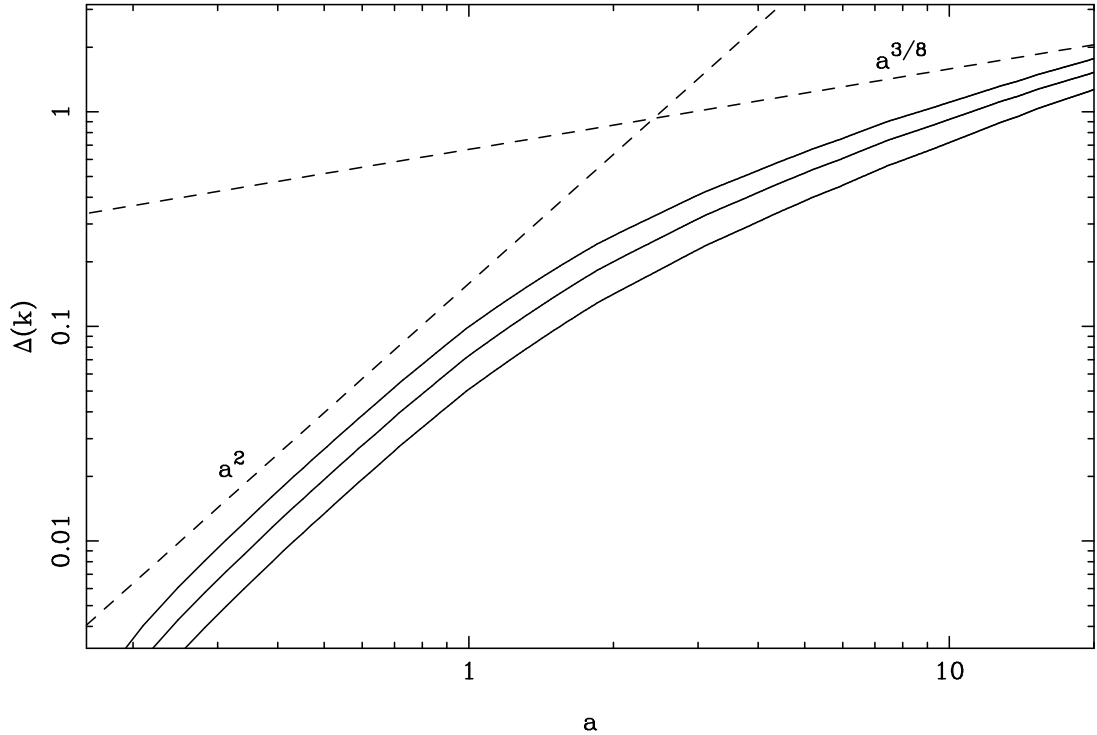}
\caption{This figure shows the rate of growth for power at some modes
in the $k^4$ tail.  For comparison we have plotted a dashed line for
growth proportional to $a^{2}$ and another for $a^{3/8}$.  This figure
clearly shows that at early times the square root of the power grows
in proportion with 
$a^2$ and at late times the rate of growth drops below $a$.  However,
it does not reach the expected quasi-linear rate till the end of this
simulation.}
\end{figure}

\subsection{Influence of Small Scales on Large Scales}

Let us now consider the flow of power to larger scales in
gravitational clustering. It is well-known that the motion of
particles conserving momentum leads to a $k^4$ tail to the power
spectrum, if the original power was sub-dominant to $k^4$ at small $k$
[\cite{k4tail}, \cite{pjep74}]. Figure~6 shows the tail for a
simulation which had initial power peaked around $k_0=2\pi/L_0 ;
L_0=8$. The initial power spectrum was a narrow Gaussian with
amplitude adjusted so that the peak reaches non-linearity at
$a=0.25$. In initial stages of evolution, there is evidence for a
$k^4$ tail. The amplitude of tail grows as $a^4$ initially, in
comparison with the linear rate $a^2$, and slows down at later
stages. If we assume the quasi-linear evolution is governed by equation
\ref{hamilton} then the index $n=4$ will change to $n=-3/8$. In
figure~6, we have plotted a line with this slope for reference. There
is some 
evidence for the slope approaching this value; the evolution of slope
is definitely in the right direction. [It is clear from lower panel of
figure~5 that the index of a $n=4$ spectrum evolves rapidly, even when
$\bar\xi < 1$. For example, the index evolves to less than $3$ for
$\bar\xi = 0.1$.] Shandarin and Melott \shortcite{k42d} have studied the
evolution of the $k^4$ tail in detail using {\it two} dimensional
numerical simulations. They also note the change of slope and a slow
decline in the rate of growth of the $k^4$ tail.

We have plotted the rate of growth for a few modes in the $k^4$ tail
as a function of scale factor.  This is shown in figure~7 that also
has two reference lines for growth proportional to $a^2$ and to
$a^{3/8}$.  The latter is the rate of growth in the
quasi-linear regime obtained from the scaling relation
eqn.(\ref{hamilton}.)  It is clear that the rate of growth for all the
modes shown here starts out as $a^2$ but drops to a rate slower than
$a$ at late times.

We stress that the evolution of power outside the band containing
initial power is entirely due to power transfer by non-linear
mode coupling. While at smaller scales this transfer is significant
and leads to equal amount of kinetic energy per logarithmic wave band,
the flow of power to larger scales is less. One can easily see that
$k^{-3/8}$ spectrum will contribute an amount of energy $k^{5/8}$ per
logarithmic band. There is less energy at larger wavelengths, i.e at
smaller $k$. It is, of course, understandable on general grounds that
large scales will not be affected by the strong non-linearities in the
small scales [see e.g., the discussion in \S{28} of Peebles
\shortcite{lssu}].

\section{Conclusions}

In conclusion, we note that the transfer of power in gravitational
clustering shows some generic pattern which is worth exploring
further. Figure~4 demonstrates that small scales [even if
highly non-linear] do not influence larger scales. The dominance of
cascading over inverse cascading
as well as the existence of a universal index for the induced the
power spectrum is reminiscent of fluid turbulence. It may be possible
to use some of the concepts from the study of turbulence to make
ideas like critical indices, fixed points, equipartition of energy,
etc. sharper and build a new paradigm for understanding non-linear
gravitational clustering.

\section*{ACKNOWLEDGMENT}

This work was completed while one of the authors (TP) was
visiting Department of Astrophysics, Princeton University supported by
NASA grant NAG5-2759. JSB thanks CSIR India for continued support. The
authors thank J.P.Ostriker and P.J.E.Peebles for useful discussions.

\label{lastpage}

\end{document}